# Observation of Fano resonance in photoluminescence of a two-dimensional magnetic semiconductor

Pingfan Gu[1,†], Qinghai Tan[2,†], Yi Wan[1,†], Ziling Li[1], Yuxuan Peng[1], Jiawei Lai[3], Junchao Ma[3], Xiaohan Yao[1], Kai Yuan[1], Dong Sun[3,4], Bo Peng[5], Jun Zhang[2,6] and Yu Ye[1,4*]

[1]State Key Laboratory for Artificial Microstructure & Mesoscopic Physics, Nano-optoelectronics Frontier Center of the Ministry of Education, School of Physics, Peking University, Beijing 100871, China.
[2]State Key Laboratory of Superlattices and Microstructures, Institute of Semiconductors & Center of Materials Science and Optoelectronics Engineering, University of Chinese Academy of Sciences, Beijing 100049, China.
[3]International Center of Quantum Matter, Beijing 100871, China.
[4]Collaborative Innovation Center of Quantum Matter, Beijing 100871, China.
[5]National Engineering Research Center of Electromagnetic Radiation Control Materials and State Key Laboratory of Electronic Thin Films and Integrated Devices, School of Microelectronics and Solid State Electronics, University of Electronic Science and Technology of China, Chengdu 610054, China.
[6]Beiing Academy of Quantum Information Science, Beijing 100193, China.
[†]These authors contributed equally to this work.
[*]Correspondence and requests for materials should be addressed to Y.Y. (email: ye_yu@pku.edu.cn).

**Quantum interference gives rise to the asymmetric Fano resonance line shape when the final states of an electronic transition follows within a continuum of states and a discrete state, which has significant applications in optical switching and sensing. The resonant optical phenomena associated with Fano resonance have been observed by absorption spectra, Raman spectra, transmission spectra,**

**etc., but have rarely been reported in photoluminescence (PL) spectroscopy. In this work, we performed spectroscopic studies on layered chromium thiophosphate (CrPS$_4$), a promising ternary antiferromagnetic semiconductor with PL in near-infrared wavelength region and observed Fano resonance when CrPS$_4$ experiences phase transition into the antiferromagnetic state below Néel temperature (38 K). The photoluminescence of the continuum states results from the *d* band transitions localized at Cr$^{3+}$ ions, while the discrete state reaches saturation at high excitation power and can be enhanced by the external magnetic field, suggesting it is formed by an impurity level from extra atomic phosphorus. Our findings provide insights into the electronic transitions of CrPS$_4$ and their connection to its intrinsic magnetic properties.**

The recently discovered two-dimensional (2D) van der Waals (vdWs) magnetic crystals are now of tremendous scientific interest, for their important role in advancing our understanding of the quantum nature of these materials and fueling the opportunities in low-power ultra-compact 2D spintronic devices. In early 2017, Ising-type and Heisenberg-type ferromagnetic orders were discovered in atomically thin CrI$_3$ and Cr$_2$Ge$_2$Te$_6$, respectively.[1, 2] In contrast to the magnetic insulators, vdWs Fe$_3$GeTe$_2$ joined the 2D magnetic family later as a ferromagnetic conductor, exhibiting gate-tunable room-temperature anomalous Hall effect.[3, 4] Due to the excellent electrostatic integrity of 2D materials and their integration of vdWs heterostructures,[5] these itinerant magnets and magnetic insulators have revealed diverse application perspectives, e.g. giant tunneling magnetoresistance,[6-11] and control of the spin and valley pseudospin.[12] However, the techniques of probing the 2D magnetism and studying its mechanism are very limited, due to its small size and magnetic momentum.

Spectroscopic studies of the light-matter interactions in 2D magnets have proven to be powerful methods for unraveling 2D magnetism,[13-17] among which photoluminescence (PL) spectroscopy is one of the most effective approaches.[16, 17] For instance, the ferromagnetic single-layer CrI$_3$ emits circularly polarized photons below the Curie temperature when excited with linearly polarized light, and the chirality of the emitted photons can be switched by the magnetization direction of the single-layer CrI$_3$.[16] However, among all the 2D magnetic materials discovered up to now, very few of them exhibit PL. Consequently, the rich photo-physics of 2D magnets, e.g. quantum interference and symmetry breaking related phenomena, has not been observed in PL spectra.

Chromium thiophosphate (CrPS$_4$) is a promising vdWs antiferromagnetic semiconductor with PL in the near-infrared region.[18, 19] The bulk CrPS$_4$ was synthesized thirty years ago and reported an antiferromagnetic order of $T_N$ of ~36 K.[20] As the vdWs gaps exist between the sulfur layers along the $c$ axis, single to hundreds of layers CrPS$_4$ flakes can be obtained using the mechanical exfoliation technique. In this work, we observed Fano resonance in the PL spectra of the layered CrPS$_4$ below its Néel temperature, which is the result of the quantum interference effect between the discrete and continuum states. The observation of Fano resonance in the 2D magnetic semiconductor provides us with a platform to gain deep insight of resonance phenomena, which is of great importance to the design of photonic devices. By means of temperature, magnetic field and excitation wavelength dependent studies, we deduced that the PL of the continuum states is derived from the $d$ band transitions localized at Cr$^{3+}$ ions, while the discrete state from the extra atomic phosphorus. Our findings not only provide insight of the electronic transitions of CrPS$_4$, but also prove that spectroscopy can serve as an important means to study the magnetic properties of 2D materials.

The crystal structure of CrPS$_4$ exhibits a monoclinic symmetry with $a$=10.871 Å, $b$=7.254 Å, $c$=6.140 Å, $\beta$=91.88°, volume=483.929 Å$^3$, space group of $C_2^3$, and $Z$=4.[21] As illustrated in Figure 1(a), puckered S layers are arranged in hexagonal close-packing parallel to $a$ axis. Two symmetrically inequivalent Cr atoms are both coordinated by six S atoms in the form of a slightly distorted octahedron. The P atoms are located in the centers of S tetrahedrons.

We synthesized single-crystal bulk CrPS$_4$ via the chemical vapor transport (CVT) method. The mixture of powdered elements of Cr, P, and S with a molar ratio of 1:1:4 and a total mass of 105 mg were sealed in an evacuated quartz ampule. The ampule was then placed in a two-zone furnace, where the source and sink temperatures for the growth were set at 750 °C and 650 °C, respectively, and kept for a week. Subsequently, the furnace was slowly cooled down to room temperature and high-quality CrPS$_4$ crystals were obtained. The bulk CrPS$_4$ can be exfoliated onto a Si/SiO$_2$ substrate by the Scotch tape method, as showed in the inset of Figure 1(c).

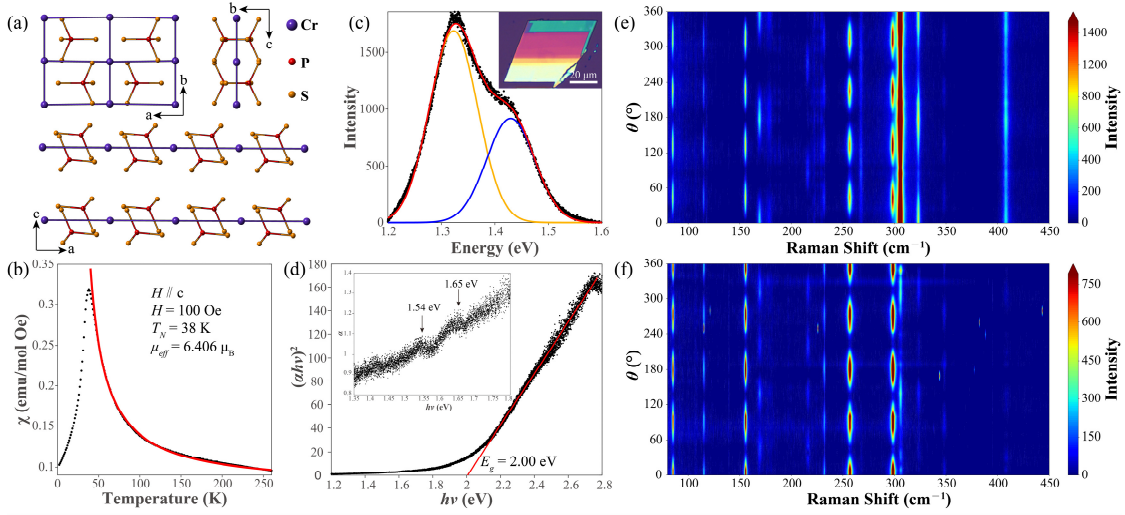

**Figure 1. Basic characterization of CrPS$_4$.** (a), The crystal structure of CrPS$_4$ viewed along *a*, *b* and *c* axes. (b), Experimental data (black dots) of the susceptibility of CrPS$_4$ single crystals on the dependence of temperature and the least-square analysis by Curie-Weiss law (red line) in paramagnetic state. (c), Experimental room-temperature PL spectrum (black dots) of the layered CrPS$_4$ and the fitting results (red line). The spectral line can be well fitted by two Gaussian profiles, located at 1.32 eV and 1.43 eV, respectively. The inset shows the layered CrPS$_4$ flake exfoliated on a Si/SiO$_2$ substrate. (d), $(\alpha h\nu)^2$ versus $h\nu$ of a few-layer CrPS$_4$ flake (black dots). Linear fitting (red line) by Equation (4) reveals the fundamental absorption edge at 2.00 eV. The inset shows the zoomed-in absorption coefficient measured by the differential reflection spectrum. Two absorption peaks located at 1.54 eV and 1.65 eV can be observed. (e-f), Polarization dependence of Raman spectra at parallel (e) configuration and vertical (f) configuration.

A series of basic characterization was performed afterward. The energy dispersive spectroscopy (EDS) implies the composition ratio of the flake is close to 1:1:4 for Cr:P:S, where phosphorus is slightly superfluous (Supporting Material Table S1). Figure 1(b) shows the temperature dependence of susceptibility ($\chi(T)$) with the applied magnetic field $H$=0.01 T along the direction of *c* axis, the easy magnetization axis. The measurement results show the Néel temperature is 38 K, consistent with the previous reports.[20] Above Néel temperature, the crystal is in paramagnetic state and the temperature dependence of $\chi$ follows the Curie-Weiss law:

$$\chi(T) = \frac{M}{H} = \frac{C}{T - \theta_P} + \chi_0 \qquad (1)$$

From the least-square analysis by Equation (1), we can obtain the Weiss temperature $\theta_P$=21.0 K and Curie constant $C$=513.2 emu·K/mol. The effective magnetic moment $\mu_{eff}$ can then be obtained by:

$$\mu_{eff} = \sqrt{C \cdot \frac{3k_B}{N_A \mu_0}} = 6.41\ \mu_B \qquad (2)$$

which is larger than that of the theoretical value of $Cr^{3+}$ ion (3.87 $\mu_B$), due to the strong spin-lattice coupling in 2D crystals.[22]

The room-temperature PL spectrum (Figure 1(c)) is composed of two emission peaks located at 1.32 eV and 1.43 eV. The intensities of these two peaks exhibit two-lobed shapes with two maximum intensity angles as the excitation or the detection polarization varies and can be well fitted by $\cos^2\theta$ function (Supporting Material Figure S3). These signatures indicate that the emission peaks may result from linearly polarized dipole transitions.[23] According to the crystal structure of $CrPS_4$, each Cr atom is located in a distorted octahedral interstice formed by six S atoms, thus inducing an octahedral ligand perturbation field that split the $3d$ orbitals of $Cr^{3+}$ into $t_{2g}$ and $e_g$ orbitals with splitting energy of $\Delta_0 \sim 2.00$ eV.[24] The two emission peaks are attributed to the lowest spin-allowed $d$-$d$ transitions of $Cr^{3+}$ ions, $t_{2g}^3(^4A_2) \rightarrow t_{2g}^2 e_g(^4T_2)$ and $t_{2g}^3(^4A_2) \rightarrow t_{2g}^2 e_g(^4T_1)$, as have been observed in many other Cr-based compounds.[25, 26] To further investigate the electronic structure, we conducted optical absorption measurements of few-layer $CrPS_4$. For a layered material with thickness $d \ll \lambda$, ($\lambda$ is the wavelength of incident light), the absorptance $A$ can be obtained by measuring the fractional change of reflectance spectrum:[27]

$$\delta_R = \frac{R_{sample} - R_{substrate}}{R_{sample}} = \frac{4}{n_0^2 - 1} A \qquad (3)$$

where $R_{sample}$ and $R_{substrate}$ are the reflectance spectra of the sample and the underlying substrate, $n_0$ is the refractive index of the substrate. The absorptance $A$ is proportional to the absorption coefficient $\alpha$ in the assumption that the absorbed intensity $dI$ is far less than the incident intensity $I_0$. From the plot of $(\alpha h\nu)^2$ versus the photon energy $h\nu$ (Figure 1(c)), we can determine the absorption edge of 2.00 eV by the following equation:[28]

$$\alpha \propto \frac{1}{h\nu}\sqrt{h\nu - E_g} \qquad (4)$$

where the $E_g$ is the bandgap of few-layer CrPS4. The obtained absorption edge is consistent with that of reflection electron energy loss spectroscopy (REELS) results,[24] which was ascribed to the charge-transfer transitions from the $3p$ of S band to unoccupied $3d$ band of Cr atom.[24] However, as the abovementioned *d* band transitions are forbidden according to Laporte parity selection rule and are weakly allowed due to local symmetry breaking, they don't exhibit obvious absorption features.[16] In the energy below the absorption edge, two weak peaks can be observed at 1.54 eV and 1.65 eV, as shown in the inset of Figure 1(d). The energy difference of these two absorption peaks precisely matches that of the emission peaks (1.32 eV and 1.43 eV). The large Stokes shift of 220 meV might result from the Franck-Condon principle and strong electron-lattice coupling,[29] which was also observed in other Cr-based compounds.[16, 17]

The Raman spectrum of the layered CrPS4 consists of 19 peaks at the excitation wavelength of 633 nm (Supporting Materials Figure S1), all of which can be well fitted using a Lorentzian line shape, revealing the complicated lattice vibration behaviors of the CrPS4 2D crystal. As shown in Figure 1(a), distances between neighboring Cr atoms along *a* axis are longer than that of *b* axis,[21] inducing a structural in-plane anisotropy. As a result, the Raman modes exhibit strong rotation angle dependence under parallel- (Figure 1(e)) and cross-polarization (Figure 1(f)) configurations in the 80-450 cm$^{-1}$ range. Some of them are assigned as A modes, showing a two-lobed shape in the parallel-polarization configuration and a four-lobed shape in the cross-polarization configuration, while others assigned as B modes, showing 90° periodic variations in both parallel- and cross-configurations (Supporting Materials Figure S2).

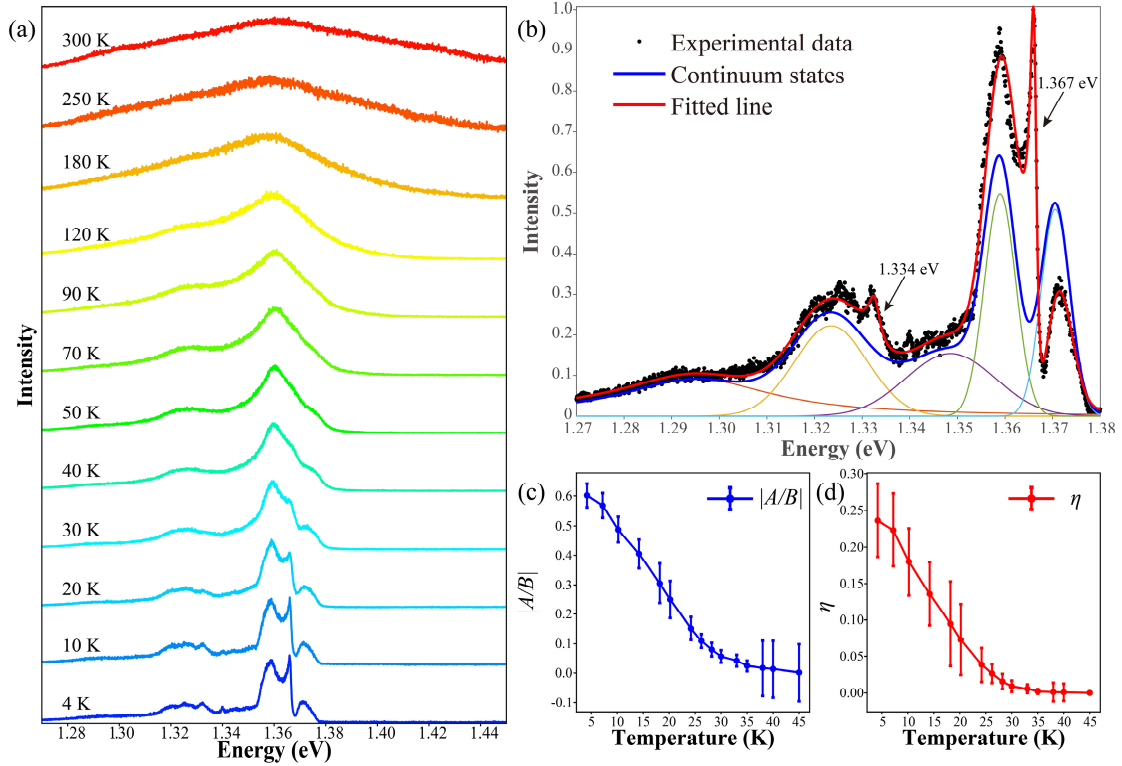

**Figure 2. Fano resonance of CrPS$_4$ on the dependence of temperature.** (a), PL spectra of CrPS$_4$ from 4 K to room temperature. (b), PL spectrum at 4 K fitted by Fano formula. The continuum band is composed of 4 Gaussian peaks and 1 Lorentzian peak. (c), The ratio of intensity between the discrete state and continuum states $|A/B|$ on the dependence of temperature. (d), The coupling coefficient of the continuum component that interferes with the discrete state $\eta$ on the dependence of temperature.

Table 1. Emission peaks at low temperature

| Peak Position(eV) | 1.296(2) | 1.324(5) | 1.333(8) | 1.349(6) | 1.360(0) | 1.367(3) | 1.371(6) |
| --- | --- | --- | --- | --- | --- | --- | --- |
| Line Shape | Lorentzian | Gaussian | Fano | Gaussian | Gaussian | Fano | Gaussian |

To establish the correlation between electronic transition and the antiferromagnetic order in CrPS$_4$, we measured the Raman (Supporting Materials Figure S1) and PL spectra (Figure 2(a)) of a layered CrPS$_4$ as a function of temperature. The Raman spectrum at low temperatures is not much different from that at room temperature (Supporting Materials Table S2), while the PL spectrum shows a significant change when the sample is gradually heated up from 4 K to room temperature. The original two peaks at room temperature split into several peaks during the cooling process, and the linewidths decrease monotonically due to the broadening of lifetimes. Remarkably, below the Néel temperature of CrPS$_4$ (38 K), two asymmetric profiles located at 1.333(8) eV and 1.367(3) eV emerge and become more salient as temperature decreases. The asymmetric line-shape is a typical feature of Fano resonance and can be observed in different samples.

In general, Fano resonance describes the quantum interference effect between the discrete and continuum states. Here, the Fano resonance is observed because the electronic transitions involve multiple channels, one of these channels possess a narrow resonance in which the spectral dependence of phase undergoes a change of π, forming the so-called discrete state, while other channels do not show a sharp change in the phase, forming the continuum states.[30] Assuming the wave function of the discrete state can be described by a Lorentzian profile $Ae^{i\delta_A}(\epsilon + i)^{-1}$ and the continuum states can be described by a constant $Be^{i\delta_B}$, the resulting emission intensity can be described by the Fano formula:[31]

$$I(\omega) = \left(\frac{(q+\Omega)^2}{1+\Omega^2}\eta + (1-\eta)\right)B^2 \tag{5}$$

where $q$ is the Fano parameter, characterizing the specific asymmetric profile of the Fano response function, $\Omega = (\omega - \omega_0)/(\Gamma/2)$ is the dimensionless frequency, $\omega_0$ is the central frequency, and $\Gamma$ is the width of the narrow band. The coupling coefficient $\eta \in [0, 1]$ is the ratio of the continuum component that interferes with the discrete state,

$$\eta = \frac{2F\cos^2\delta}{F + 2\sin\delta + \sqrt{F^2 + 4F\sin\delta + 4}} \tag{6}$$

where $F = A/B$ and $\delta = \delta_A - \delta_B$ are the relative intensity ratio and phase difference between the discrete state and continuum states.

Considering the mechanism of Fano resonance, the Fano parameter $q$ is also responsible for the ratio of the transition probabilities to the effective discrete state $\Phi$ and the continuum state $\psi_E$ in the form:

$$\frac{1}{2}\pi q^2 = \frac{|\langle\Phi|T|i\rangle|^2}{|\langle\Psi_E|T|i\rangle|^2}\Gamma$$

According to the above discussion, the three parameters in Equation (6) (7) $q$, $\eta$, $F$ must not be independent. They are correlated in the form:

$$q = \frac{F\cos\delta}{\eta} \qquad (8)$$

We used the above equations to fit the normalized PL spectra at different temperatures, with the most salient feature at 1.367 eV being the main focus of our study. $F$ and $\eta$ are set as two independent variables to describe the emission intensity through different channels on the dependence of temperature. As shown in Figure 2(b), the continuum band is composed of 5 emission peaks. 5 linearly summed Voigt profiles are used to determine the composition of the emission peaks,[32] and the line shapes and the peak positions of them are summarized in Table 1. The linewidth analysis shows that four peaks are pure Gaussian components, where the bandwidth broadening is dominated by the dephasing process. Only the peak at the lowest energy can be fitted into a Lorentzian function, whose bandwidth is contributed by natural broadening. Assuming the phase difference $\delta$ to be a constant, the temperature dependent PL signature of Fano resonance can be described by two crucial parameters, the ratio of intensity between the discrete state and continuum states $|A/B|$ and the coupling coefficient $\eta$. As shown in Figure 2(c), the coupling coefficient, which is maximum at the lowest temperature, exhibits drastic decrease as temperature rises and approaches 0 at a temperature above 35 K, indicating the continuum states no longer couple with the discrete state above the Néel temperature of CrPS$_4$. The parameter $|A/B|$ experiences a similar decrease (Figure 2(d)), indicating the decay of the discrete state contribution in the emission processes. The disappearance of Fano resonance above the Néel temperature of CrPS$_4$ can either be caused by the disappearance of the discrete state or by the decoupling of the continuum and discrete states. Since no sharp emission peak is observed at 1.367 eV at temperatures above 35 K, the disappearance of the Fano resonance can only be attributable to the fact that the discrete state is completely inhibited above the antiferromagnetic phase transition temperature.

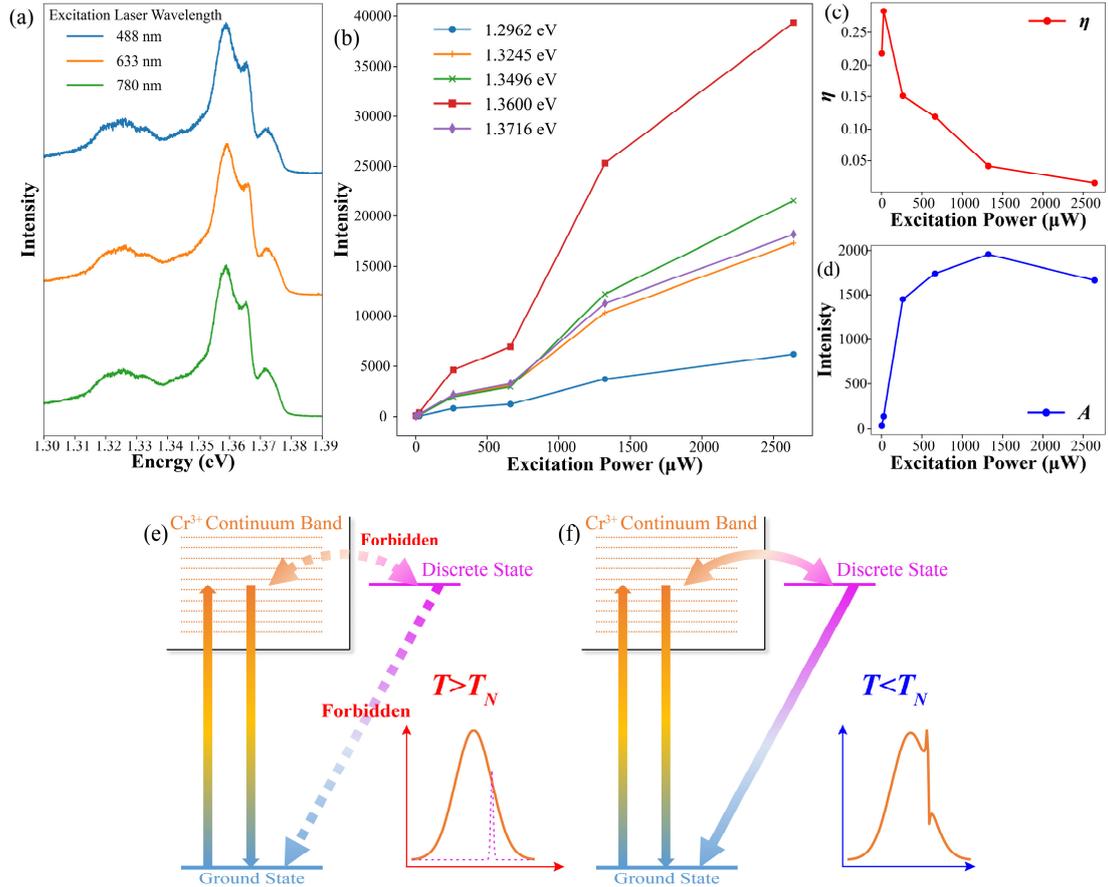

**Figure 3. Mechanism of PL Fano resonance in CrPS$_4$.** (a), Fano resonance can be observed under excitation with different wavelengths. (b), Peak intensities of the five fitted continuum states on the dependence of incident excitation power. (c), The coupling coefficient $\eta$ on the dependence of incident excitation power. (d), Peak intensity of the discrete state on the dependence of incident excitation power. (e-f), Mechanism of Fano resonance in CrPS$_4$. (e), Above Néel temperature of CrPS$_4$, the discrete state transition is forbidden, thus only continuum states transitions are observed. (f), Below Néel temperature, the antiferromagnetic order of CrPS$_4$ breaks the time reversal symmetry, which makes the forbidden discrete state transition occur and interfere with the continuum states, leading to a Fano-type resonance.

To probe the origin of the continuum states and the discrete state interfering with each other, we studied the Fano resonance of CrPS$_4$ with varying excitation wavelength and power. Under three different excitations (488 nm, 633 nm, and 780 nm), the asymmetric line-shape can all be observed at the same energy of 1.367(3) eV (Figure 3(a)), ruling out the possibilities of Raman scattering or Fabry-Pérot interference related phenomena. Under the 780 nm laser excitation, we studied incident laser power dependent Fano resonance of CrPS$_4$ at 4 K in detail (Supporting

Materials Figure S4). The PL intensities of all 5 continuum sates scale linearly with the excitation power and exhibit no trend of saturation at high power (Figure 3(b)), suggesting the PL of the continuum states arise from the intrinsic electronic transitions of CrPS$_4$. In addition, the peak positions of the continuum states do not vary with temperature (Figure 2(b)) and the PL peaks at 4 K exhibit identical excitation anisotropy to the peaks at room temperature (Supporting Materials Figure S5). Therefore, it is reasonable to infer that the PL of the continuum states at low temperature also results from the orbital transitions of the Cr$^{3+}$ ions, identical to those at room temperature. Distinct from the continuum states, the PL intensity of the discrete state (extracted from $A = F \cdot B|_{E=1.3673 \text{ eV}}$), saturates rapidly as the laser power increases (Figure 3(d)). The limited density of states suggests that the discrete state PL results from an extrinsic energy level, e.g., induced by defect or impurity.[33] As the incident power increases, the PL intensity of the continuum states increases linearly while that of the discrete state saturates, leading to the decrease of coupling coefficient $\eta$ (Figure 3(c)) and obscuring the Fano resonance at high excitation power (Supporting Materials Figure S4).

The energy of Fano resonance, 1.367 eV, is in accord with the energy difference between the two excitation states of a neutral P atom, $^2$D$_{5/2}$(Ne$3s^23p^24p$) and $^4$P$_{3/2}$(Ne$3s^23p^34s$).[34] As mentioned above, phosphorus is slightly superfluous in our CrPS$_4$ crystal. As a result, the extra atomic phosphorus is supposed to form a discrete energy level, to which the transition plays the role of the discrete state that interferes with transitions to the continuum states in Fano resonance. Distinct from the previously reported Fano resonance, the Fano resonance in CrPS$_4$ is observed in PL spectra, so the actual quantum interference happens during the photon emission process. Above the Néel temperature of CrPS$_4$, the transition of the discrete energy level is forbidden due to the conservation of spin angular momentum, thus only PL from continuum states transitions is observed (Figure 3(e)). Below the Néel temperature of CrPS$_4$, the antiferromagnetic order of CrPS$_4$ breaks the time reversal symmetry, which makes the forbidden discrete state transition occur and interfere with the continuum states, leading to a Fano-type asymmetric line shape (Figure 3(f)).

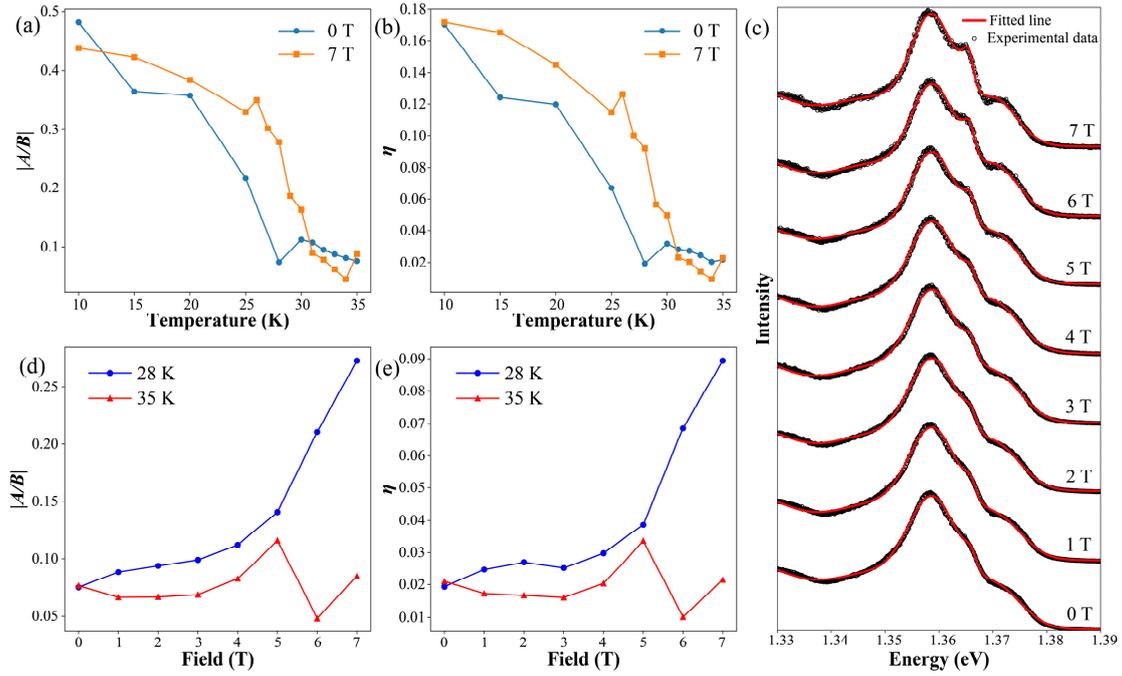

**Figure 4. The influence of the magnetic field to Fano resonance.** (a-b), The Fano parameter $|A/B|$ (a) and $\eta$ (b) on the dependence of the temperature under the applied magnetic field of 0 T and 7 T. (c), PL spectra of CrPS$_4$ at 28 K under different applied field. From 0 T to 7 T, the Fano resonance is gradually enhanced. (d-e), The Fano parameter $|A/B|$ (e) and $\eta$ (f) on the dependence of magnetic field from 0 T to 7 T at 28 K and 35 K.

We can see that the antiferromagnetic order of CrPS$_4$ is of great importance to the occurrence of Fano resonance. It is reasonable to envision that an external magnetic field may also have a similar effect on enabling the forbidden discrete transition,[35] either by flipping the localized spin or by breaking time-reversal symmetry.[36] Figure 4(a) and 4(b) compare the temperature-dependent Fano parameters $|A/B|$ and $\eta$ under external magnetic field of 0 T and 7 T. In the application field of 7 T, these two parameters show clear increase from 10 K to 30 K, and the Fano resonance disappearing temperature increases by 5 K, indicating the Fano resonance is indeed enhanced (Supporting Materials Figure S5). The PL spectra of CrPS$_4$ at 28 K show strong magnetic field dependence (Figure 4(c)). The Fano parameters $|A/B|$ (Figure 4(d)) and $\eta$ (Figure 4(e)) are almost constant at the low field but experience rapid increases above 4 T. The reoccurrence of Fano resonance under high magnetic field indicates the selection rule is broken and the discrete state transition emerges. However, the origin of the observed magnetic field threshold and the microscopic explanation of the applied field interacting with the Fano resonance require further investigation. At the temperature below 10 K (above 35 K), the applied field shows no obvious influence on the PL spectra (Figure 4(d) and 4(e)), because of the saturation of the discrete state transition due to the intrinsic antiferromagnetic order of $CrPS_4$ (the disappearance of the antiferromagnetic order of $CrPS_4$).

In summary, we observed Fano resonance in layered antiferromagnetic semiconductor $CrPS_4$. This is the first Fano resonance observed in the intrinsic PL spectrum of the material. By studying the dependence of Fano resonance on temperature, magnetic field, laser polarization, etc., we infer that the continuum states are resulted from the *d-d* transitions of $Cr^{3+}$ ions, and the discrete state is formed by the extra atomic phosphorus. Below Néel temperature, the antiferromagnetic order of $CrPS_4$ results in broken symmetry and the primarily forbidden discrete state transition emerges, resulting in Fano resonance. Our findings not only provide insight of the electronic transitions of $CrPS_4$, but also prove that spectroscopy can serve as an important means to study the magnetic properties of 2D materials.

## Acknowledgements

This work was supported by the National Key R&D Program of China (Grant Nos. 2017YFA0206301, 2018YFA0306900 and 2017YFA0303401), Beijing Natural Science Foundation (JQ18014), Strategic Priority Research Program of Chinese Academy of Sciences (grant no. XDB28000000). We thank S.J. for the help in sealing the ampules.

## Supporting Information

Supplementary Information accompanies this paper.

## Conflict of Interest

The authors declare no competing financial interests.

## Keywords

# References

[1]   B. Huang, G. Clark, E. Navarro-Moratalla, D. R. Klein, R. Cheng, K. L. Seyler, D. Zhong, E. Schmidgall, M. A. McGuire, D. H. Cobden, W. Yao, D. Xiao, P. Jarillo-Herrero, X. Xu, *Nature* **2017**, *546*, 270.

[2]   C. Gong, L. Li, Z. Li, H. Ji, A. Stern, Y. Xia, T. Cao, W. Bao, C. Wang, Y. Wang, Z. Q. Qiu, R. J. Cava, S. G. Louie, J. Xia, X. Zhang, *Nature* **2017**, *546*, 265.

[3]   Z. Fei, B. Huang, P. Malinowski, W. Wang, T. Song, J. Sanchez, W. Yao, D. Xiao, X. Zhu, A. F. May, W. Wu, D. H. Cobden, J. H. Chu, X. Xu, *Nat. Mater.* **2018**, *17*, 778.

[4]   Y. Deng, Y. Yu, Y. Song, J. Zhang, N. Z. Wang, Z. Sun, Y. Yi, Y. Z. Wu, S. Wu, J. Zhu, J. Wang, X. H. Chen, Y. Zhang, *Nature* **2018**, *563*, 94.

[5]   C. Gong, X. Zhang, *Science* **2019**, *363*, eaav4450.

[6]   Z. Wang, I. Gutierrez-Lezama, N. Ubrig, M. Kroner, M. Gibertini, T. Taniguchi, K. Watanabe, A. Imamoglu, E. Giannini, A. F. Morpurgo, *Nat. Commun.* **2018**, *9*, 2516.

[7]   T. Song, X. Cai, M. W. Tu, X. Zhang, B. Huang, N. P. Wilson, K. L. Seyler, L. Zhu, T. Taniguchi, K. Watanabe, M. A. McGuire, D. H. Cobden, D. Xiao, W. Yao, X. Xu, *Science* **2018**, *360*, 1214.

[8]   D. R. Klein, D. MacNeill, J. L. Lado, D. Soriano, E. Navarro-Moratalla, K. Watanabe, T. Taniguchi, S. Manni, P. Canfield, J. Fernandez-Rossier, P. Jarillo-Herrero, *Science* **2018**, *360*, 1218.

[9]   H. H. Kim, B. Yang, T. Patel, F. Sfigakis, C. Li, S. Tian, H. Lei, A. W. Tsen, *Nano Lett.* **2018**, *18*, 4885.

[10]  D. Ghazaryan, M. T. Greenaway, Z. Wang, V. H. Guarochico-Moreira, I. J. Vera-Marun, J. Yin, Y. Liao, S. V. Morozov, O. Kristanovski, A. I. Lichtenstein, M. I. Katsnelson, F. Withers, A. Mishchenko, L. Eaves, A. K. Geim, K. S. Novoselov, A. Misra, *Nat. Electron.* **2018**, *1*, 344.

[11]  Z. Wang, D. Sapkota, T. Taniguchi, K. Watanabe, D. Mandrus, A. F. Morpurgo, *Nano Lett.* **2018**, *18*, 4303.

[12]  D. Zhong, K. L. Seyler, X. Linpeng, R. Cheng, N. Sivadas, B. Huang, E. Schmidgall, T. Taniguchi, K. Watanabe, M. A. McGuire, W. Yao, D. Xiao, K. C. Fu, X. Xu, *Sci. Adv.* **2017**, *3*, e1603113.

[13]  J. U. Lee, S. Lee, J. H. Ryoo, S. Kang, T. Y. Kim, P. Kim, C. H. Park, J. G. Park, H. Cheong, *Nano Lett.* **2016**, *16*, 7433.

[14]  X. Wang, K. Du, Y. Y. Fredrik Liu, P. Hu, J. Zhang, Q. Zhang, M. H. S. Owen, X. Lu, C. K. Gan, P. Sengupta, C. Kloc, Q. Xiong, *2D Mater.* **2016**, *3*, 031009.

[15]  Y. Tian, M. J. Gray, H. Ji, R. J. Cava, K. S. Burch, *2D Mater.* **2016**, *3*, 025035.

[16]  K. L. Seyler, D. Zhong, D. R. Klein, S. Gao, X. Zhang, B. Huang, E. Navarro-Moratalla, L. Yang, D. H. Cobden, M. A. McGuire, W. Yao, D. Xiao, P. Jarillo-Herrero, X. Xu, *Nat. Phys.* **2017**, *14*, 277.

[17]  Z. Zhang, J. Shang, C. Jiang, A. Rasmita, W. Gao, T. Yu, *Nano Lett.* **2019**, *19*, 3138.

[18]  Q. L. Pei, X. Luo, G. T. Lin, J. Y. Song, L. Hu, Y. M. Zou, L. Yu, W. Tong, W. H. Song, W. J. Lu, Y. P. Sun, *J. Appl. Phys.* **2016**, *119*, 043902.

[19]  J. Lee, T. Y. Ko, J. H. Kim, H. Bark, B. Kang, S. G. Jung, T. Park, Z. Lee, S. Ryu, C. Lee, *ACS Nano* **2017**, *11*, 10935.

[20]  R. Berc, D. M. Schleich, G. Ouvrard, A. Louisy, J. Rouxel, *Inorg. Chem.* **1979**, *18*, 1814.

[21]  R. Diehl, C. D. Carpentier, *Acta Crystallogr. Sect. B* **1977**, *33*, 1399.

[22]  M. A. McGuire, G. Clark, S. Kc, W. M. Chance, G. E. Jellison, V. R. Cooper, X. Xu, B. C. Sales, *Phys. Rev. Mater.* **2017**, *1*, 014001.

[23]  T. T. Tran, K. Bray, M. J. Ford, M. Toth, I. Aharonovich, *Nat. Nanotechnol.* **2016**, *11*, 37.

[24]  Y. Ohno, A. Mineo, I. I. Matsubara, *Phys. Rev. B: Condens. Matter* **1989**, *40*, 10262.

[25]  D. L. Wood, J. Ferguson, K. Knox, J. F. Dillon, *J. Chem. Phys.* **1963**, *39*, 890.

[26]  L. Nosenzo, G. Samoggia, I. Pollini, *Phys. Rev. B: Condens. Matter* **1984**, *29*,

3607.

[27] K. F. Mak, M. Y. Sfeir, Y. Wu, C. H. Lui, J. A. Misewich, T. F. Heinz, *Phys. Rev. Lett.* **2008**, *101*, 196405.

[28] M. S. Dresselhaus, *Solid state physics part ii optical properties of solids*, 2001.

[29] J. Garacia Sole, L. E. Bausa, D. Jaque, *An introduction to the optical spectroscopy of inorganic solids*, John Wiley & Sons, 2005.

[30] U. Fano, *Phys. Rev.* **1961**, *124*, 1866.

[31] M. V. Rybin, P. V. Kapitanova, D. S. Filonov, A. P. Slobozhanyuk, P. A. Belov, Y. S. Kivshar, M. F. Limonov, *Phys. Rev. B: Condens. Matter* **2013**, *88*, 205106.

[32] B. Sontheimer, M. Braun, N. Nikolay, N. Sadzak, I. Aharonovich, O. Benson, *Phys. Rev. B: Condens. Matter* **2017**, *96*, 121202.

[33] T. Schmidt, K. Lischka, W. Zulehner, *Phys. Rev. B: Condens. Matter* **1992**, *45*, 8989.

[34] W. C. Martin, R. Zalubas, A. Musgrove, *J. Phys. Chem. Ref. Data* **1985**, *14*, 751.

[35] V. Bellani, E. Perez, S. Zimmermann, L. Vina, *Solid State Commun.* **1996**, *97*, 459.

[36] X. X. Zhang, T. Cao, Z. Lu, Y. C. Lin, F. Zhang, Y. Wang, Z. Li, J. C. Hone, J. A. Robinson, D. Smirnov, S. G. Louie, T. F. Heinz, *Nat. Nanotechnol.* **2017**, *12*, 883.